\documentclass[a4paper,8pt]{article}

\usepackage{latexsym}
\usepackage{a4wide}
\usepackage{graphicx, subfigure} 
\usepackage{cite} 
\usepackage{tabularx}
\usepackage{array}
\usepackage{graphics}     
\usepackage{amsmath}
\usepackage{amssymb}
\usepackage{longtable} 
\usepackage{verbatim} 
\usepackage[table]{xcolor}
\usepackage{booktabs}
\usepackage{hyperref} 
\usepackage[utf8]{inputenc}
\usepackage{multirow}
\usepackage{soul}

\definecolor{light-gray}{gray}{0.90}

\newcommand{\cen}[1]{\multicolumn{1}{c}{#1}}

\begin{document}

\hfill {\tt  CERN-TH-2017-110, IPM/P.A-466, MITP/17-035}  

\def\thefootnote{\fnsymbol{footnote}}
 
\begin{center}

\vspace{3.cm}

{\Large\bf {On lepton non-universality in exclusive $b\to s \ell\ell$ decays} }

\setlength{\textwidth}{11cm}
                    
\vspace{2.cm}
{\large\bf  
T.~Hurth\footnote{Email: tobias.hurth@cern.ch}$^{,a}$,
F.~Mahmoudi\footnote{Also Institut Universitaire de France, 103 boulevard Saint-Michel, 75005 Paris, France; Email: nazila@cern.ch}$^{,b,c,d}$,
D.~Mart\'inez Santos\footnote{Email: Diego.Martinez.Santos@cern.ch}$^{,e}$,
S.~Neshatpour\footnote{Email: neshatpour@ipm.ir }$^{,f}$
}
 
\vspace{1.cm}
{\em $^a$PRISMA Cluster of Excellence and  Institute for Physics (THEP)\\
Johannes Gutenberg University, D-55099 Mainz, Germany}\\[0.2cm]
{\em $^b$Univ Lyon, Univ Lyon 1, ENS de Lyon, CNRS, Centre de Recherche Astrophysique de Lyon UMR5574, F-69230 Saint-Genis-Laval, France}\\[0.2cm]
{\em $^c$Univ Lyon, Univ Lyon 1, CNRS/IN2P3, Institut de Physique Nucl\'eaire de Lyon UMR5822,\\ F-69622 Villeurbanne, France}\\[0.2cm]
{\em $^d$Theoretical Physics Department, CERN, CH-1211 Geneva 23, Switzerland}\\[0.2cm]
{\em $^e$Instituto Galego de F\'isica de Altas Enerx\'ias, Universidade de Santiago de Compotela, Spain}\\[0.2cm]
{\em $^f$School of Particles and Accelerators,
Institute for Research in Fundamental Sciences (IPM)
P.O. Box 19395-5531, Tehran, Iran}

\end{center}

\renewcommand{\thefootnote}{\arabic{footnote}}
\setcounter{footnote}{0}

\vspace{1.cm}
\thispagestyle{empty}
\centerline{\bf ABSTRACT}
\vspace{0.5cm}
The LHCb measurements of certain ratios of decay modes testing lepton flavour non-universality might open an exciting world of new physics beyond the standard model. The latest LHCb measurements of $R_{K^*}$ offer some new insight beyond the previous measurement of $R_K$. We work out the present significance for non-universality, and argue that claims of $5\sigma$ deviations from the Standard Model based on all present $b \to s \ell^+\ell^-$  data including the ratios are misleading and are at present still based on guesstimates of hadronic power corrections in the $b \to s \ell^+\ell^-$ angular observables.

We demonstrate that only a small part of the luminosity of 50 fb$^{-1}$ foreseen to be accumulated by the LHCb will be needed to offer soon a definite answer to the present question of whether we see a very small glimpse of lepton flavour non-universal new physics or not. 

We also present new predictions for other ratios based on our analysis of the present measurements of the ratios  $R_{K^{(*)}}$ and analyse if they are able to differentiate between various new physics options within the effective field theory at present or in the near future.

\newpage
\section{Introduction}
The ratio $R_{K^*} \equiv \Gamma(B \to K^{*} \mu^{+} \mu^{-})\,/\,\Gamma(B \to K^{*} e^{+} e^{-})$ has been recently measured by the LHCb \mbox{collaboration} in two bins of the dilepton mass~\cite{Aaij:2017vbb} reporting deviations of $2.2-2.4$  and $2.4-2.5\sigma$ from the Standard Model (SM) respectively:
\begin{equation}  \label{RKstar}
R_{K^*}( [0.045,\,1.1]{\rm GeV^2}) = 0.660^{+0.110}_{-0.070}\pm0.024, \hspace{1cm} R_{K^*} ([1.1,\,6.0] {\rm GeV^2}) = 0.685^{+0.113}_{-0.069}\pm0.047\,,
\end{equation}
where the first errors are statistical and the second systematic.  This measurement establishes another hint for lepton flavour non-universality besides the previously measured ratio $R_K \equiv \Gamma(B^{+}\to K^{+} \mu^{+} \mu^{-})\,/\,\Gamma(B^{+}\to K^{+} e^{+} e^{-})$~\cite{Aaij:2014ora} which represents a $2.6 \sigma$ deviation from the SM prediction 
\begin{equation}  \label{eq:RK}
R_K([1,\,6]{\rm GeV^2}) = 0.745 ^{+ 0.090}_{-0.074}\pm 0.036  \,.
\end{equation}
These observables are theoretically clean  because hadronic uncertainties cancel out in the ratios
~\cite{Hiller:2003js}. One could think that electromagnetic corrections, in particular logarithmically enhanced QED corrections, might play a role in these observables in a lepton non-universal way. However, these corrections of the form $\alpha\, {\rm log^2}(m_b/m_\ell)$ were calculated in the inclusive case and were shown to be rather well simulated by the PHOTOS Monte Carlo which is also used by the LHCb collaboration~\cite{Huber:2015sra,Huber:2007vv}. More recently, these corrections were directly estimated in the exclusive case~\cite{Bordone:2016gaq} confirming this conclusion. Our SM predictions for these three observables based on SuperIso v3.7~\cite{Mahmoudi:2007vz,Mahmoudi:2008tp} are the following:
\begin{equation}
R_{K^*}([0.045,1.1]) =0.906 \pm 0.022\,,\quad   R_{K^*}( [1.1,\,6] ) =0.997 \pm 0.01\,,\quad      R_{K} ([1,\,6]) = 1.000\pm 0.01\,.       
\label{eq:SM}
\end{equation}
We have taken over the analysis of QED corrections from Ref.~\cite{Bordone:2016gaq}. The 
larger error in the very low bin is due to larger QED corrections, $2\%$, and due to the input parameters, $1\%$. The latter error is dominated by form factor uncertainties which do not fully cancel out.\footnote{In our case the error due to the input parameters (dominantly due to form factors) is much smaller than the one quoted in Refs.~\cite{Bordone:2016gaq,Capdevila:2017bsm}. We think in the case of Ref.~\cite{Capdevila:2017bsm} this is due to the difference that  we use the full form factor calculation  presented in Ref.~\cite{Straub:2015ica} while the authors of Ref.~\cite{Capdevila:2017bsm} use the results~\cite{Khodjamirian:2005ea,Khodjamirian:2010vf} which are based on another LCSR method. This method has much larger uncertainties somehow by construction. 
We have further analysed the dependence of the theory error of $R_{K^*}$ in the very low bin on the form factor error.
We tripled the error  given in the  LCSR calculation of Ref.~\cite{Straub:2015ica} and found a $1.3\%$ error due to the form factor (and other input parameters only) in the prediction of $R_{K^*}$ in the  low bin. 
In any case this feature will become relevant only in the future when the statistical error is reduced.
We state that we use a Monte Carlo analysis where all the input parameters as well as the involved scales and form factors are varied randomly, taking into account all the correlations.}  
Presently, the statistical errors of the experimental measurements are the dominating ones in all three observables.

We confirm the SM deviations claimed by the LHCb collaboration for each of the three measurements. We obtain $2.3 \sigma$ ($R_{K^*}$ low), $2.5 \sigma$ ($R_{K^*}$ central) and $2.6 \sigma$ ($R_{K}$ central), respectively.
By combining the three measurements we arrive at a SM deviation of $3.6 \sigma$. 

Beyond these theoretically clean flavour observables there are many other SM deviations  in the present $b\to s \ell \ell$
data, in particular in the angular observables of the $B\to K^* \mu\mu$ decays~\cite{Aaij:2015oid} and in the branching ratios of the $B_s \to \phi \mu\mu$ decay~\cite{Aaij:2015esa}. However, as emphasised in our previous analyses~\cite{Hurth:2013ssa,Hurth:2014vma,Hurth:2016fbr,Mahmoudi:2016mgr,Chobanova:2017ghn},
all these observables are affected by unknown (non-factorisable) power corrections which can only be {\it guesstimated} at present.
In contrast to the claim in Ref.~\cite{Capdevila:2017bsm} this issue is {\it not} resolved but  makes it rather difficult or even impossible to separate new physics (NP) effects from such potentially large hadronic power corrections within these exclusive angular observables and branching ratios.  As a  consequence,  the significance of these deviations depends on the  assumptions made within such a guesstimate of the unknown power corrections. This is of course also true if these observables are combined with the theoretically clean ratios $R_K$ and $R_{K^*}$ within a global analysis of all $b\to s\ell\ell$ data. In this sense, claims that such global analyses indicate a large deviation from the SM above the $5\sigma$ level are misleading as long as the precise assumptions made on the non-factorisable power corrections are not clearly indicated. Hence a real estimate of the non-factorisable power corrections is highly desirable to disentangle NP effects from hadronic uncertainties in the angular observables (see Sec.~\ref{sec:nonfactorizable}).

However, the present tensions in  $R_K$ on the one side and in the angular observables in $B \to K^*\mu\mu$ and branching ratios  in $B_s \to \phi \mu\mu$ on the other side can be explained  by a similar NP contribution $C_9$ to the semileptonic operator as was demonstrated in various global analyses (see for example~\cite{Hurth:2014vma,Altmannshofer:2014rta,Descotes-Genon:2015uva,Hurth:2016fbr}). We analyse this question including the measurements of $R_{K^*}$ in the following. Such a coherent picture -- if found -- is an exciting and strong result which indicates that the NP interpretation is a valid option for the explanation of the tensions in the angular observables, but it should not be misinterpreted as a proof for the NP option at present. But such a feature also implies that  a confirmation of the deviations in the ratios would indirectly confirm the NP interpretation of the anomalies in the angular observables in $B \to K^*\mu\mu$. 

These findings suggest a separate analysis of the theoretically clean ratios $R_K$ and $R_{K^*}$ and similar ratios of this kind testing lepton universality, which we present in the following -- based on our previous analyses in Refs.\cite{Hurth:2016fbr,Hurth:2014vma} (see Sec.~\ref{sec:combined}). Only in a second step, we compare our findings with a global fit to all the $b\to s \ell\ell$ data excluding the ratios (Sec.~\ref{sec:comparison}).

Moreover, we analyse the prospects of the LHCb collaboration to give a definite answer to the question of whether the present deviations from the SM predictions represent a very small glimpse of lepton flavour non-universal new physics or not (Sec.~\ref{sec:future}).

We will see that the present deviations within  the three ratios can be explained by NP contributions to six different Wilson coefficients. We analyse predictions of many other ratios which are sensitive to possible lepton flavour non-universal new physics in order to examine the possibility of distinguishing between the different new physics options (Sec.~\ref{sec:predictions}).

\begin{table}
\begin{center}
\setlength\extrarowheight{3pt}
\scalebox{0.90}{
\begin{tabular}{|c|ccc|}\hline
 & \cen{b.f. value} & $\chi^2_{\rm min}$ & ${\rm Pull}_{\rm SM}$  \\ 
\hline \hline
$\Delta C_{9}$                                          & $-0.48$ & $18.3$  & $0.3\sigma$   \\ 
$\Delta C_{9}^{\prime}$                                 & $+0.78$ & $18.1$ & $0.6\sigma$   \\ 
$\Delta C_{10}$                                         & $-1.02$ & $18.2$ & $0.5\sigma$  \\ 
$\Delta C_{10}^{\prime}$                                & $+1.18$ & $17.9$ & $0.7\sigma$ \\ 
$\Delta C_{9}^{\mu}$                                    & $-0.35$ & $5.1$  & $3.6\sigma$     \\ 
$\Delta C_{9}^{e}$                                      & $+0.37$ & $3.5$  & $3.9\sigma$     \\ 
\multirow{ 2}{*}{$\Delta C_{10}^{\mu}$}           & $-1.66 $ & \multirow{ 2}{*}{$2.7$}  & \multirow{ 2}{*}{$4.0\sigma$} \vspace{-0.1cm} \\
                                                  & $-0.34 $ &                          &     \\ 
\multirow{ 2}{*}{$\Delta C_{10}^{e}$}             & $-2.36 $ & \multirow{ 2}{*}{$2.2$}  & \multirow{ 2}{*}{$4.0\sigma$}  \vspace{-0.1cm} \\
                                                  & $+0.35 $ &                          &     \\ \hline
\end{tabular} \quad \quad \quad 
\begin{tabular}{|c|ccc|}\hline 
 & \cen{b.f. value} & $\chi^2_{\rm min}$ & ${\rm Pull}_{\rm SM}$  \\ 
\hline \hline 
$\Delta C_{9}^{\mu}=-\Delta C_{10}^{\mu}$ ($\Delta C_{\rm LL}^\mu$)               & $-0.16$ & $3.4$  & $3.9\sigma$   \\ 
$\Delta C_{9}^{e}=-\Delta C_{10}^{e}$ ($\Delta C_{\rm LL}^e$)                     & $+0.19$ & $2.8$  & $4.0\sigma$   \\ 
$\Delta C_{9}^{\mu \prime}=-\Delta C_{10}^{\mu \prime}$ ($\Delta C_{\rm RL}^\mu$) & $-0.01$ & $18.3$ & $0.4\sigma$   \\ 
$\Delta C_{9}^{e \prime}=-\Delta C_{10}^{e \prime}$ ($\Delta C_{\rm RL}^e$)       & $+0.01$ & $18.3$ & $0.4\sigma$   \\ 
$\Delta C_{9}^{\mu}=+\Delta C_{10}^{\mu}$ ($\Delta C_{\rm LR}^\mu$)                & $+0.09$ & $17.5$ & $1.0\sigma$   \\ 
$\Delta C_{9}^{e}=+\Delta C_{10}^{e}$ ($\Delta C_{\rm LR}^e$)                      & $-0.55$ & $1.4$  & $4.1\sigma$  \\ 
$\Delta C_{9}^{\mu \prime}=+\Delta C_{10}^{\mu \prime}$ ($\Delta C_{\rm RR}^\mu$)  & $-0.01$ & $18.4$ & $0.2\sigma$   \\ 
$\Delta C_{9}^{e \prime}=+\Delta C_{10}^{e \prime}$ ($\Delta C_{\rm RR}^e$)        & $+0.61$ & $2.0$  & $4.1\sigma$   \\ \hline  
\end{tabular}}
\caption{Best fit values in the one-operator fits (where only one Wilson coefficient is varied at a time) considering \emph{only} the observables ${R_{K^*}}_{[0.045,1.1]},{R_{K^*}}_{[1.1,6]}$ and ${R_K}_{[1,6]}$. 
The $\delta C_i$ in the fits are normalised to their SM values according to $\Delta C_{i}^{(\prime)} \equiv \delta C_{i}^{(\prime)}/C_{i}^{\rm SM}$
with $C_9^{\rm SM}=4.20$ and $C_{10}^{\rm SM}=-4.01$, 
and in the right table the normalisation is always with $C_9^{\rm SM}$. When two numbers are mentioned for a given $\Delta C_{i}$, they correspond to two possible minima.
\label{tab:FLVObs}} 
\end{center} 
\end{table}

\vspace*{0.3cm} 
\section{Combined analysis of the $R_K^{(*)}$ ratios} \label{sec:combined}
The tensions of the measurements of these three ratios with the SM predictions can be explained in a model-independent way by modified Wilson coefficients ($C_i= C_i^{SM} + \delta C_i$), where $\delta C_i$ can  be due to some NP  effects. First we consider the impact of NP in one Wilson coefficient at a time, where all other Wilson coefficients are kept to their SM values.
Assuming such a scenario to be the correct description of the three ratios, the SM value for the Wilson coefficient $C_i$  (corresponding to $\delta C_i=0$) is in a specific tension with the best fit value (${\rm Pull}_{\rm SM}$).
In Table~\ref{tab:FLVObs} we give SM pulls of the various one-operator hypotheses.~\footnote{Regarding the notations, we recall that in Ref.~\cite{Alonso:2014csa} the semi-leptonic operators ${\cal O}_9^{(')}$ and ${\cal O}_{10}^{(')}$ within the electroweak Hamiltonian were singled out as the only operators which can explain 
the deviation in the ratio $R_K$:
\begin{equation}
{\cal O}_9^{(')\ell}= (\alpha_{\rm em}/4\pi)     (\bar{s}    \gamma_\mu P_{L(R)} b)\;(\bar{\ell}  \gamma^\mu\ell)\,, \quad
{\cal O}_{10}^{(')\ell}= (\alpha_{\rm em}/ 4\pi)  (\bar{s} \gamma_\mu P_{L(R)} b)\;(\bar{\ell}   \gamma^\mu \gamma_5  \ell)\,.
\end{equation}
In order to account for lepton non-universality, one considers separate electron and muon semi-leptonic operators, 
$\ell = \mu,\, e $.  
The corresponding Wilson coefficients are denoted as $C_{9,10}^{(')\, e}$ and $C_{9,10}^{(')\,\mu}$ respectively which are equal in the SM or in models with lepton universality. Under  the assumption that there are left-handed leptons only -- which represents an attractive option in model building beyond the SM -- one finds the following relations between the Wilson coefficients:
\begin{equation}
\delta C_{LL}^{\ell}\equiv \delta C_9^{\ell} =- \delta C_{10}^{\ell}\,,\,\,\,\,\,  \delta C_{RL}^{\ell} \equiv \delta C^{' \,\ell}_9 = -\delta C^{'\,\ell}_{10}\,.
\end{equation}
Here we introduced the quantities $C_{XY}^{\ell}$ where $\ell$ is again the flavour index, $X$ denotes the chirality of the quark current and $Y$ of the lepton current. Assuming right-handed leptons only, one gets the following relations: 
\begin{equation}
\delta C_{RR}^{\ell} \equiv \delta C^{' \,\ell}_9 = + \delta C^{' \,\ell}_{10}\,,\,\,\,   \delta C_{LR}^{\ell} \equiv \delta{C^{\ell}_9} =  + \delta C^{\ell}_{10}\,.
\end{equation}
}
We see that NP in {$C_9^e$, $C_9^\mu$}, $C_{10}^e$, or $C_{10}^\mu$ are favoured by the $R_{K^{(*)}}$ ratios
with a significance of $3.6 - 4.0 \sigma$. NP contributions in primed operators have no significant effect in a better description of the data.
Among the chiral Wilson coefficients, we find four with a SM pull around $3.9 - 4.1 \sigma$, namely $C_{LL}^\mu$, $C_{LL}^e$, $C_{LR}^e$, and $C_{RR}^e$. The two latter ones, however, lead to a very large NP shift in the Wilson coefficient. We do not consider them in the following. Thus, there are six favoured NP one-operator hypotheses to account for the deviations in the measured ratios $R_{K^{(*)}}$. 

We present {in addition} fits based on some two-operator hypotheses (see Figure~\ref{fig:2operator} below). 
Our results are in agreement with the recent fit results presented in Refs.~\cite{Altmannshofer:2017yso,DAmico:2017mtc,Capdevila:2017bsm,Geng:2017svp,Ciuchini:2017mik,Hiller:2017bzc}.
\vspace*{0.3cm}
\begin{figure}[!h]
\begin{center}
\includegraphics[width=0.33\textwidth]{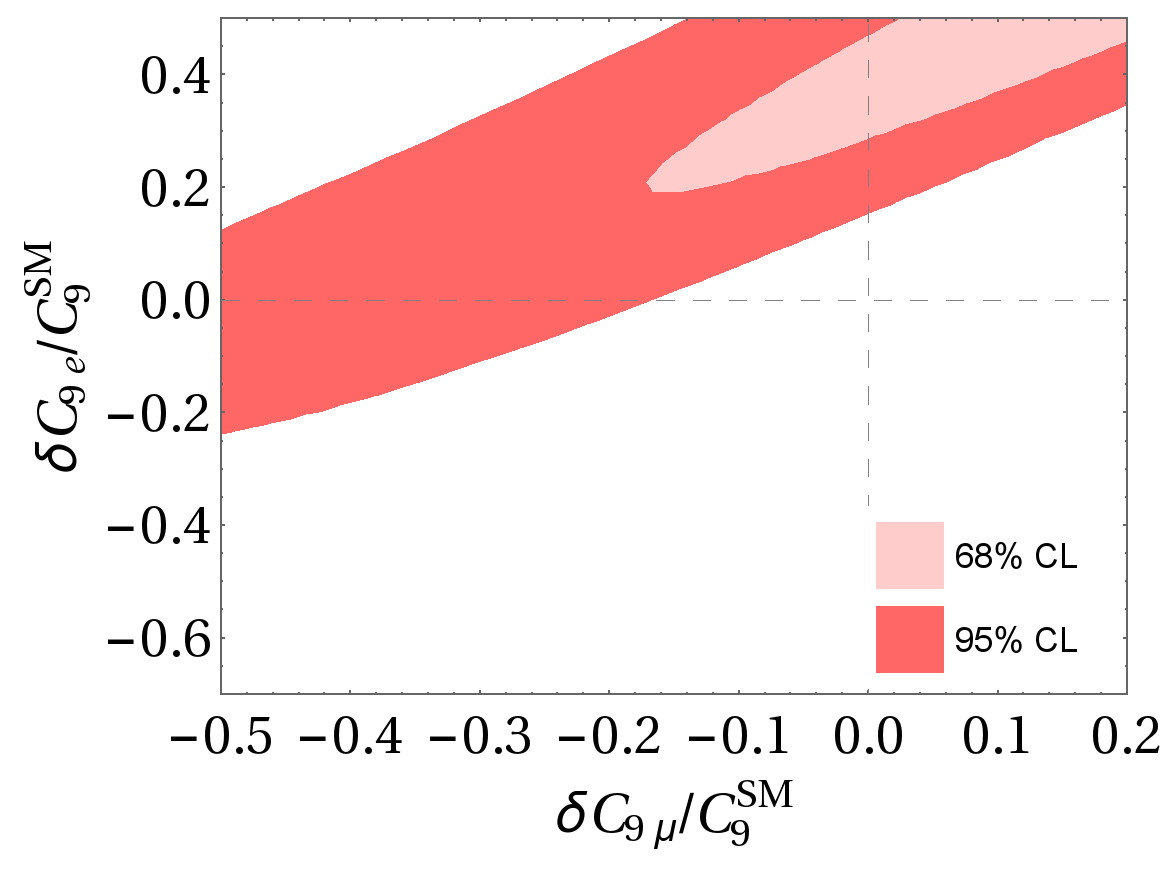}\qquad\includegraphics[width=0.33\textwidth]{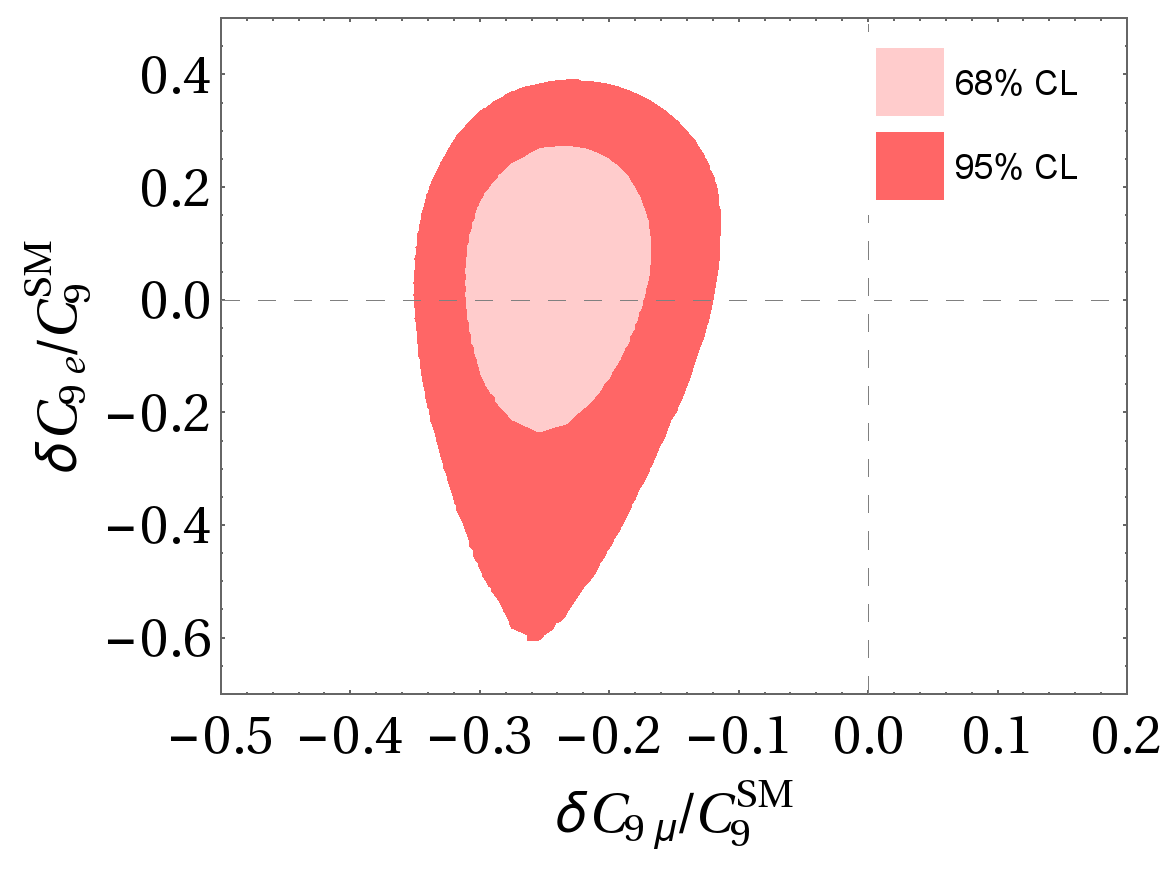}
\\[2.mm]
\includegraphics[width=0.33\textwidth]{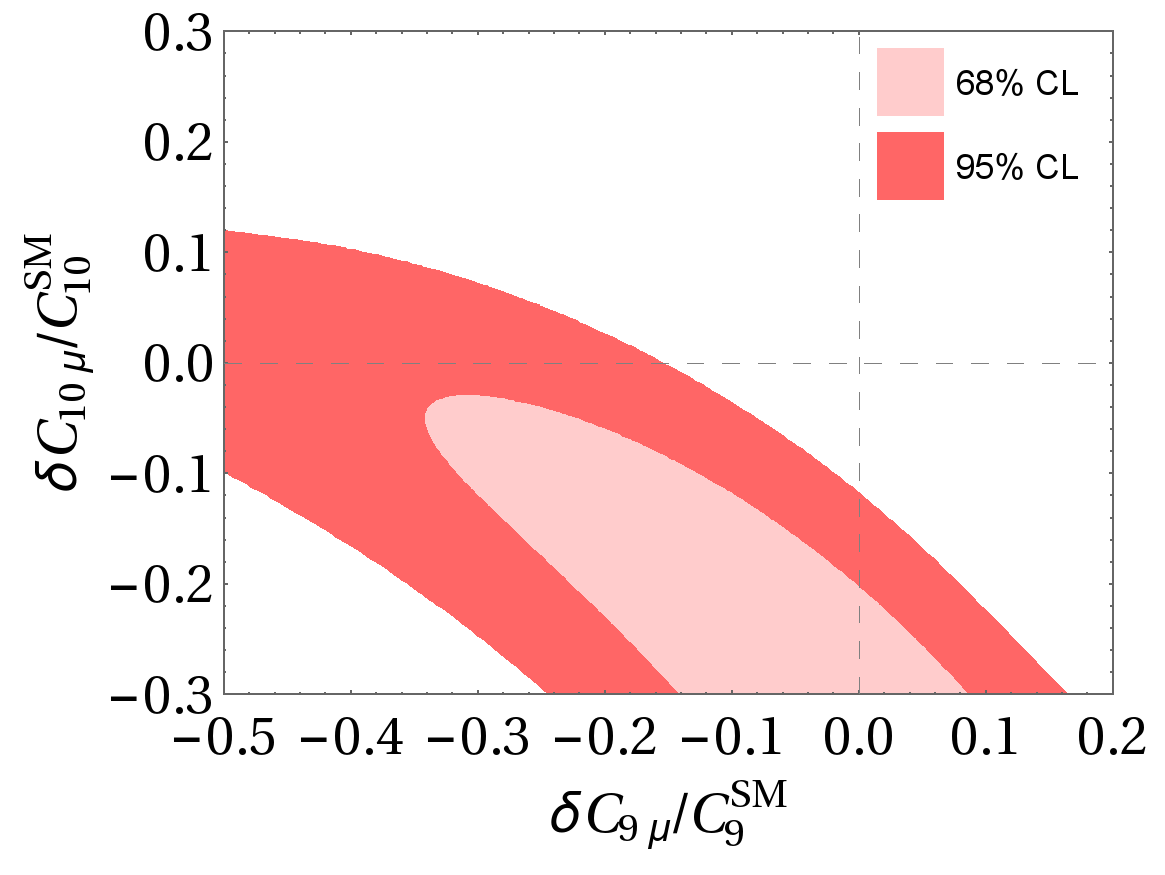}\qquad\includegraphics[width=0.33\textwidth]{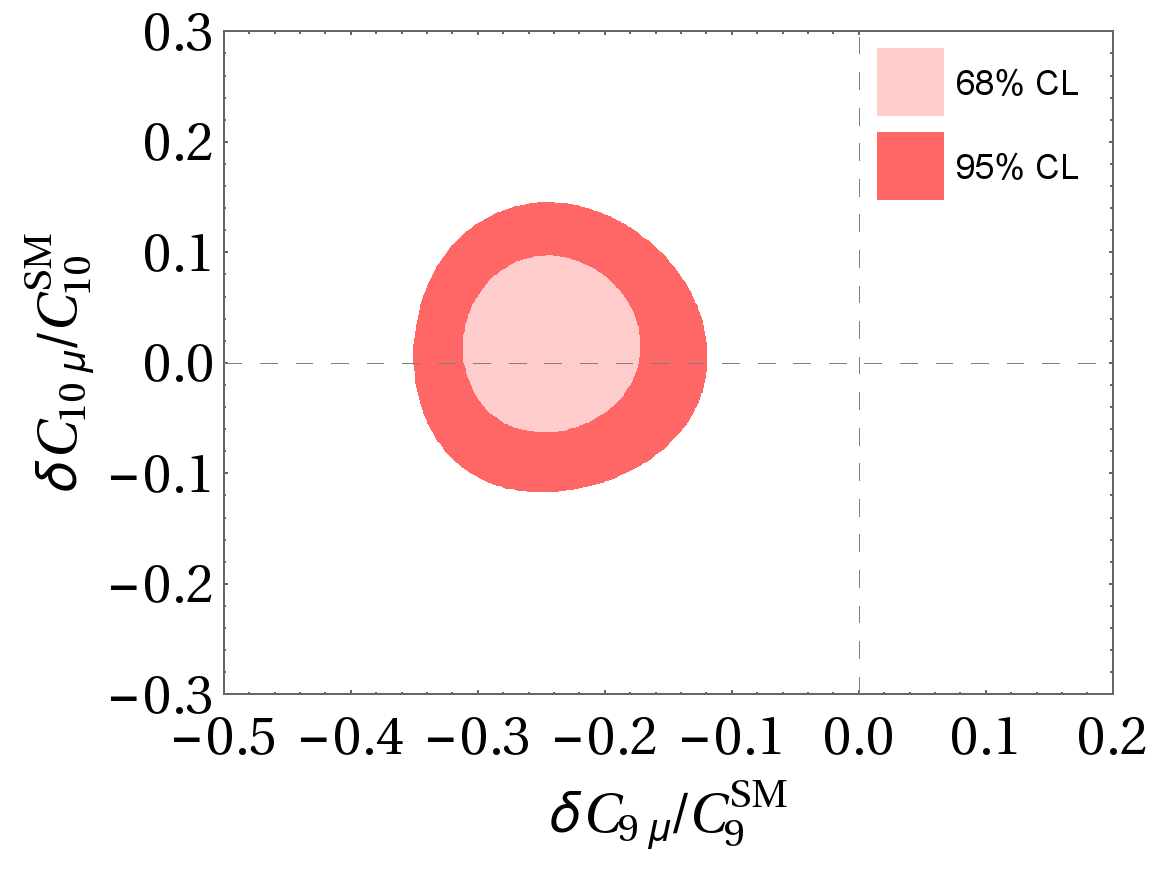}

\caption{Global fit results with present data, using only $R_K$ and $R_K^*$ (left), using all observables except $R_K$ and $R_K^*$ (under the assumption of $10\%$ non-factorisable power corrections) (right).
\label{fig:2operator}}
\end{center}
\end{figure}

\section{Comparison with the global fit excluding $R_K$ and $R_{K^*}$} \label{sec:comparison}

We redo the same exercise for all available $b \to s \ell\ell$ data without {the three $R_K^{(*)}$} ratios (for a list of the used observables, see Appendix A of Ref.~\cite{Hurth:2016fbr}). 
The SM pulls  are given in Table~\ref{tab:AllbesidesFLVObs}. We assume $10 \%$ non-factorisable power corrections. The implementation of these  corrections is done in the same way as in our previous analysis by  multiplying the hadronic terms in the QCD factorisation (QCDf) formula~\cite{Beneke:2001at,Beneke:2004dp}  which remain after putting the Wilson coefficients 
$C_{7,9,10}^{(')}$ to zero (see Sec. 3.2 of Ref.~\cite{Hurth:2016fbr}). Note that this part of the leading amplitude represents in many cases not more than one third of the complete leading QCDf amplitude, so in these cases the 
non-factorisable power corrections represent a $3\% - 4 \%$ correction to the complete amplitude only.

\begin{table}
\begin{center}
\setlength\extrarowheight{3pt}
\scalebox{0.90}{
\begin{tabular}{|c|ccc|}\hline
 & \cen{b.f. value} & $\chi^2_{\rm min}$ & ${\rm Pull}_{\rm SM}$  \\ 
\hline \hline
$\Delta C_{9}$                                          & $-0.24$ & $70.5$  & $4.1\sigma$   \\ 
$\Delta C_{9}^{\prime}$                                 & $-0.02$ & $87.4$ & $0.3\sigma$   \\ 
$\Delta C_{10}$                                          & $-0.02$ & $87.3$ & $0.4\sigma$   \\ 
$\Delta C_{10}^{\prime}$                                 & $+0.03$ & $87.0$ & $0.7\sigma$   \\ 
$\Delta C_{9}^{\mu}$                                    & $-0.25$ & $68.2$  & $4.4\sigma$     \\ 
$\Delta C_{9}^{e}$                                      & $+0.18$ & $86.2$  & $1.2\sigma$     \\ 
$\Delta C_{10}^{\mu}$                                    & $-0.05$ & $86.8$  & $0.8\sigma$      \\ 
\multirow{ 2}{*}{$\Delta C_{10}^{e}$}                    & $-2.14 $ & \multirow{ 2}{*}{$86.3$}  & \multirow{ 2}{*}{$1.1\sigma$}  \vspace{-0.1cm} \\
                                                         & $+0.14 $ &  &     \\ \hline
\end{tabular}\quad \quad \quad
\begin{tabular}{|c|ccc|} \hline
 & \cen{b.f. value} & $\chi^2_{\rm min}$ & ${\rm Pull}_{\rm SM}$  \\ 
\hline \hline
$\Delta C_{9}^{\mu}=-\Delta C_{10}^{\mu}$ ($\Delta C_{\rm LL}^\mu$)               & $-0.10$ & $79.4$  & $2.8\sigma$   \\ 
$\Delta C_{9}^{e}=-\Delta C_{10}^{e}$ ($\Delta C_{\rm LL}^e$)                     & $+0.08$ & $86.3$ & $1.1\sigma$   \\ 
$\Delta C_{9}^{\mu \prime}=-\Delta C_{10}^{\mu \prime}$ ($\Delta C_{\rm RL}^\mu$) & $-0.01$ & $87.3$ & $0.4\sigma$   \\ 
$\Delta C_{9}^{e \prime}=-\Delta C_{10}^{e \prime}$ ($\Delta C_{\rm RL}^{e}$)     & $-0.01$ & $87.0$ & $0.7\sigma$   \\ 
$\Delta C_{9}^{\mu}=+\Delta C_{10}^{\mu}$ ($\Delta C_{\rm LR}^\mu$)                & $-0.12$ & $79.5$ & $2.8\sigma$   \\ 
\multirow{ 2}{*}{$\Delta C_{9}^{e}=+\Delta C_{10}^{e }$ ($\Delta C_{\rm LR}^e$)}   & $+0.50$ & $85.8$ & $1.3\sigma$  \vspace{-0.1cm} \\ 
                                                                                  & $-1.12$ & $86.7$ & $0.9\sigma$   \\ 
$\Delta C_{9}^{\mu \prime}=+\Delta C_{10}^{\mu \prime}$ ($\Delta C_{\rm RR}^\mu$)  & $+0.03$ & $87.1$ & $0.6\sigma$   \\ 
$\Delta C_{9}^{e \prime}=+\Delta C_{10}^{e \prime}$ ($\Delta C_{\rm RR}^e$)        & $-0.54$ & $86.3$ & $1.1\sigma$    \\ \hline
\end{tabular}}
\caption{Best fit values in the one-operator fit considering all observables (under the assumption of $10\%$ factorisable power corrections) except $R_K$ and $R_{K^*}$. Normalisation is as in Table\,\ref{tab:FLVObs}. 
\label{tab:AllbesidesFLVObs}} 
\end{center} 
\end{table}

Comparing the two cases, given in Tables~\ref{tab:FLVObs} and~\ref{tab:AllbesidesFLVObs}, 
one can make some interesting observations that among the one-operator hypotheses, 
the $C_9^\mu$ solutions are favoured with SM pulls of $3.6$ and $4.4\sigma$ in the two separate fits respectively but $C_9^e$ is much less favoured in the fit to all $b \to s\ell\ell$ observables without  the ratios.  
The reason for this is that there is essentially only the branching ratio of the inclusive decay $B\to X_s e^+ e^-$  in the electron sector
within this new fit using all $b\to s\ell\ell$ observables except $R_K$ and $R_{K^*}$, and it has a much smaller deviation from the SM than the large number of muonic channels.
Primed operators have a very small SM pull in both cases; but more importantly the $C_{10}$-like solutions do not play a role in the global fit excluding the ratios 
in contrast to the $R_{K^{(*)}}$ analysis, also the chiral one-operator hypothesis $C_{LL}^\mu$ has 
less significance in comparison with the $R_{K^{(*)}}$ case. 
Thus, the NP analyses of the two sets of observables are less coherent than often stated. 
But if we consider two-operator NP hypotheses (see Figure~\ref{fig:2operator}) $(C_9^\mu,\, C_9^e)$ and $(C_{10}^\mu,\, C_9^\mu)$ 
one finds that the two sets are compatible at least at the $2\sigma$ level.\footnote{The branching ratio BR($B_s\to \mu^+ \mu^-$) is a theoretically rather clean observable.
There is good agreement between the SM prediction and the current experimental measurement of BR($B_s\to \mu^+ \mu^-$). It is well known that  this observable constrains $C_{10}$ and one might expect  the fit to the ratios changes if 
this observable is included.  However, as was shown in Tables II and III of Ref.~\cite{Geng:2017svp}, the significance for the various NP options within the one-operator hypotheses changes only very mildly when 
including BR($B_s\to \mu^+ \mu^-$)  (see also Table~\ref{tab:upgrade+Bsmumu} below).}

In this context we emphasise that the present high significance of NP effects in  $C_{10}^\mu$ within the analysis of the ratios  is not only due to the  measurement on the very low bin of $R_{K^*}$. Removing this bin, Pull$_{\rm SM}$ for 
$C_{10}^\mu$ gets only slightly reduced from 4.0$\sigma$ in Table~\ref{tab:FLVObs} (where this bin is included)
to 3.7$\sigma$ with the best fit points of $\Delta C_{10}^\mu$  changing negligibly
from -0.34 (-1.66) to -0.31 (-1.69).

\vspace*{0.5cm} 
\section{Non-factorisable power corrections}\label{sec:nonfactorizable}
Until now we worked out the global fit of all the $b \to s \ell\ell$ data under the assumption that the non-factorisable power corrections do not exceed $10\%$. But this is only a guesstimate at present. 
It was already demonstrated by several groups~\cite{Ciuchini:2015qxb,Chobanova:2017ghn} that the anomalies in the $b \to s \ell \ell$ data (without the $R_{K^{(*)}}$ ratios) can be {\it fully} explained by large non-factorisable  power corrections: The unknown non-factorisable power corrections are just fitted to the data using an ansatz with 18 real parameters. This fit  to the data needs very large non-factorisable power corrections in the critical bins -- up  to  $50\%$ or more relative to the leading QCDf amplitude. Clearly, the existence of such large power  corrections cannot be ruled out in principle and the situation stays undecided if the anomalies in this set of $b\to s\ell\ell$ observables originate  from new physics or from large hadronic power corrections. 

There are methods offered in Refs.~\cite{Khodjamirian:2010vf,Khodjamirian:2012rm} (see also Refs.~\cite{Dimou:2012un,Lyon:2013gba}) which may allow one to replace the present guesstimates of the non-factorisable power corrections by real estimates of these hadronic effects. Obviously, such estimates are highly desirable to disentangle NP effects from hadronic uncertainties in the angular observables. More recently, a slightly different approach was proposed in which the non-factorisable corrections are estimated using the analyticity structure of the corresponding amplitudes~\cite{Bobeth:2017vxj}. 
But also new experimental data on these angular observables might help to disentangle non-factorisable power corrections from new physics effects  within the angular observables. If the electron modes of the angular observables are measured and do not show any deviation, then this is a clear hint for the NP option. This is of course equivalent to a clear NP signal within the corresponding theoretically clean $R$ ratios as already discussed in the introduction.

In addition, we mention that underestimated uncertainties in the form factor determination can also at least partially account for the tensions in the global analysis of all the present $b \to s$ data (excluding the three ratios) as we have shown in Ref.~\cite{Hurth:2016fbr}.

If we now combine the two sets of observables -- assuming $10\%$ non-factorisable power corrections as this is often used as the standard choice in global analyses -- we find a SM pull of  $5.7 \sigma$ for the one-operator NP hypothesis 
$C_9^\mu$. However, NP claims based on this result are misleading. As explained, the significance of the SM pull of such a combined fit also depends directly on the guesstimates of the non-factorisable power corrections. 
These findings prevent us from making further combined fits  with  the theoretically clean ratios and  the rest of the $b \to s \ell\ell$ data.

\vspace*{0.5cm}
\section{Future prospects} \label{sec:future}
The LHCb detector will be upgraded and is expected to collect a total integrated luminosity of 50 fb$^{-1}$. 
A second upgrade at a high-luminosity LHC will allow for a full data set of up to 300 fb$^{-1}$.
Due to the expected luminosity of 300 fb$^{-1}$, of 50 fb$^{-1}$, and in the near future of 12 fb$^{-1}$ the statistical error will be decreased by a factor 10, 4, and 2, respectively.\footnote{The 12 fb$^{-1}$ is an effective luminosity, corresponding to 1 fb$^{-1}$ at 7 TeV, 2 fb$^{-1}$ at 8 TeV and 5 fb$^{-1}$ at 13 TeV.}

For the three luminosity cases we consider three upgrade scenarios in which the current central values are assumed to remain and in which the systematic error is either unchanged or reduced by a factor of 2 or 3. In all cases we consider two (extreme) options regarding the error correlations, namely that the three $R_K$ and $R_{K^*}$ bins/observables have no correlation or 50\% correlation between each of the three measurements.

The results for these future scenarios are given in Table~\ref{tab:upgrade}. Here we show the one-operator NP hypothesis $\Delta C^\mu_9$ as an exemplary mode. It is obvious from the SM pulls that -- within the  scenario in which the central values are assumed to remain --  only a small part of the 50 fb$^{-1}$ is needed to establish NP in the $R_{K^{(*)}}$  ratios even in the pessimistic case that  the systematic errors are not reduced by then at all. 

In addition we have found that the SM pulls for the six favoured one-operator NP hypotheses are all very similar in each of the upgrade scenarios. This feature can also be read off from the analytical dependence of the ratios on the NP Wilson coefficients (see Ref.~\cite{DAmico:2017mtc}). This indicates that also in future scenarios based on much larger data sets there is no differentiation between the NP hypotheses possible.  This motivates the search for other ratios in the next section which are sensitive for lepton non-universality {\it and} serve this purpose.
%
\begin{table}[t!]
\centering
\setlength\extrarowheight{2pt}
\scalebox{1.0}{
\begin{tabular}{|l||c|c|c|}\hline
\multirow{ 2}{*}{$\Delta C_9^\mu$}&   Syst.\;      & Syst./2 \;           & Syst./3  \\
                    &   Pull$_{\rm SM}$  &  Pull$_{\rm SM}$   &  Pull$_{\rm SM}$  \\  
\hline \hline
12 fb$^{-1}$   &  $6.1\sigma \;   (4.3\sigma)$  & $7.2\sigma  \; (5.2\sigma)$  & $7.4\sigma  \; (5.5\sigma)$        \\ 
50 fb$^{-1}$   & $8.2\sigma \;   (5.7\sigma)$  & $11.6\sigma  \; (8.7\sigma)$  & $12.9\sigma  \; (9.9\sigma)$        \\ 
300 fb$^{-1}$  &  $9.4\sigma \;   (6.5\sigma)$  & $15.6\sigma  \; (12.3\sigma)$  & $19.5\sigma  \; (16.1\sigma)$        \\ \hline
\end{tabular}}
\vspace*{0.1cm}
\caption{
Pull$_{\rm SM}$ for the fit to $\Delta C_9^\mu$ based on the ratios $R_K$ and $R_{K^*}$ for
the  LHCb upgrade scenarios with $12, 50$ and 300 fb$^{-1}$ luminosity collected, assuming current central values remain.
For each of the upgraded luminosities the systematic error (denoted by ``Syst.'' in the table) is considered to either remain unchanged or
be reduced by a factor of 2 or 3. In each scenario the three $R_K$ and $R_{K^*}$ bins/observables are assumed to 
have no correlation (50\% correlation between each of the three measurements).
\label{tab:upgrade}} 
\end{table}\\

If in addition to $R_K$ and $R_{K^*}$ we add the prospected measurement for the rather clean BR($B_s \to \mu^+ \mu^-$) observable into the fit, we still find that the different favoured NP scenarios cannot be differentiated. 
In Table~\ref{tab:upgrade+Bsmumu} we give Pull$_{\rm SM}$ for NP in $\Delta C_9^\mu$, $\Delta C_{10}^\mu$ or $\Delta C_{LL}^\mu$ 
considering the prospects for 
the LHCb upgrade of the $R_K$ and $R_K^*$ ratios assuming the current central values remain and 
compare it with the case when BR($B_s\to \mu^+ \mu^-$) is also included.
For the statistical errors of  $R_K$ and $R_{K^*}$ ratios  we consider the most optimistic scenario of Table~\ref{tab:upgrade}  where  
they are 1/3 of their current values without any correlations and for BR($B_s\to \mu^+ \mu^-$) we assume 5\% theoretical uncertainty with 
(combined statistical and systematic) experimental errors  
to be $0.38\times 10^{-9}$, $0.32\times 10^{-9}$, $0.26\times 10^{-9}$ for the LHCb results with 12, 50 and 300$^{-1}$ luminosity combined with the 
prospected ATLAS and CMS measurements.
The Pull$_{\rm SM}$ remains the same for the $\Delta C_9^\mu$ scenario for the $R_K$ and $R_{K^*}$ fit prospects whether BR($B_s\to \mu^+ \mu^-$) is included or not as the latter observable is insensitive to $C_9$ (the fit was redone for $\Delta C_9^\mu$ as a validation test). But also the $\Delta C_{10}^\mu$ and $\Delta C_{LL}^\mu$ change only very mildly when the prospected measurement for the rather clean BR($B_s \to \mu^+ \mu^-$) observable is added to the fit.\\
\\
%
\begin{table}[t!]
\begin{center}
\setlength\extrarowheight{3pt}
\scalebox{1.0}{
\begin{tabular}{|c||c|c|c||}
\hline 
 & \multicolumn{3}{c||}{Pull$_{\rm SM}$ with $R_K$ and $R{_K^*} \;[\;+\; {\rm BR}(B_s\to \mu^+ \mu^-)]$ prospects} \\ 
\hline 
 LHCb lum.		&   12 fb$^{-1}$     &   50 fb$^{-1}$	  &  300 fb$^{-1}$  \\
\hline
\hline
%
$C_{9}^{\mu}$ 		 & $ \phantom{B}	7.4\sigma \;\; [7.4\sigma]	\phantom{B} $ &  $ \phantom{B}	12.9\sigma \;\; [12.9\sigma]	\phantom{B} $ &  $ 	19.5\sigma \;\; [19.5\sigma]	 $  \\
$C_{10}^{\mu}$	 	 & $ 	8.1\sigma \;\; [7.6\sigma]	 $ &  $ 	13.9\sigma \;\; [13.5\sigma]	 $ &  $ 	20.8\sigma \;\; [20.6\sigma]	 $  \\
$C_{LL}^{\mu}$	 	 & $ 	7.9\sigma \;\; [7.8\sigma]	 $ &  $ 	13.6\sigma \;\; [13.6\sigma]	 $ &  $ 	20.5\sigma \;\; [20.4\sigma]	 $  \\
\hline
\end{tabular}
}\vspace*{0.1cm}
\caption{Predictions of Pull$_{\rm SM}$ for the fit to $\Delta C_9^\mu$, $\Delta C_{10}^\mu$ and $\Delta C_{LL}^\mu$ based on the ratios $R_K$ and $R_{K^*}$ [and also BR($B_s\to \mu^+ \mu^-$)] for the LHCb upgrade scenarios with $12, 50$ and 300 fb$^{-1}$ luminosity collected, assuming current central values remain.
For $R_K$ and $R_{K^*}$ in each of the upgraded luminosities we have assumed the optimistic scenario where systematic errors are reduced by a factor 3 with no correlation among the errors. For BR($B_s \to \mu^+ \mu^-$) we have considered the absolute experimental error to be
$3.8\times 10^{-10}$, $3.2\times 10^{-10}$, $2.6\times 10^{-10}$ from the prospected LHCb results with 12, 50 and 300$^{-1}$ luminosity as well as the prospected ATLAS and CMS results.\vspace*{0.8cm}
\label{tab:upgrade+Bsmumu}}
\end{center} 
\end{table}
%
\begin{figure}[!t]
\begin{center}
\includegraphics[width=0.47\textwidth]{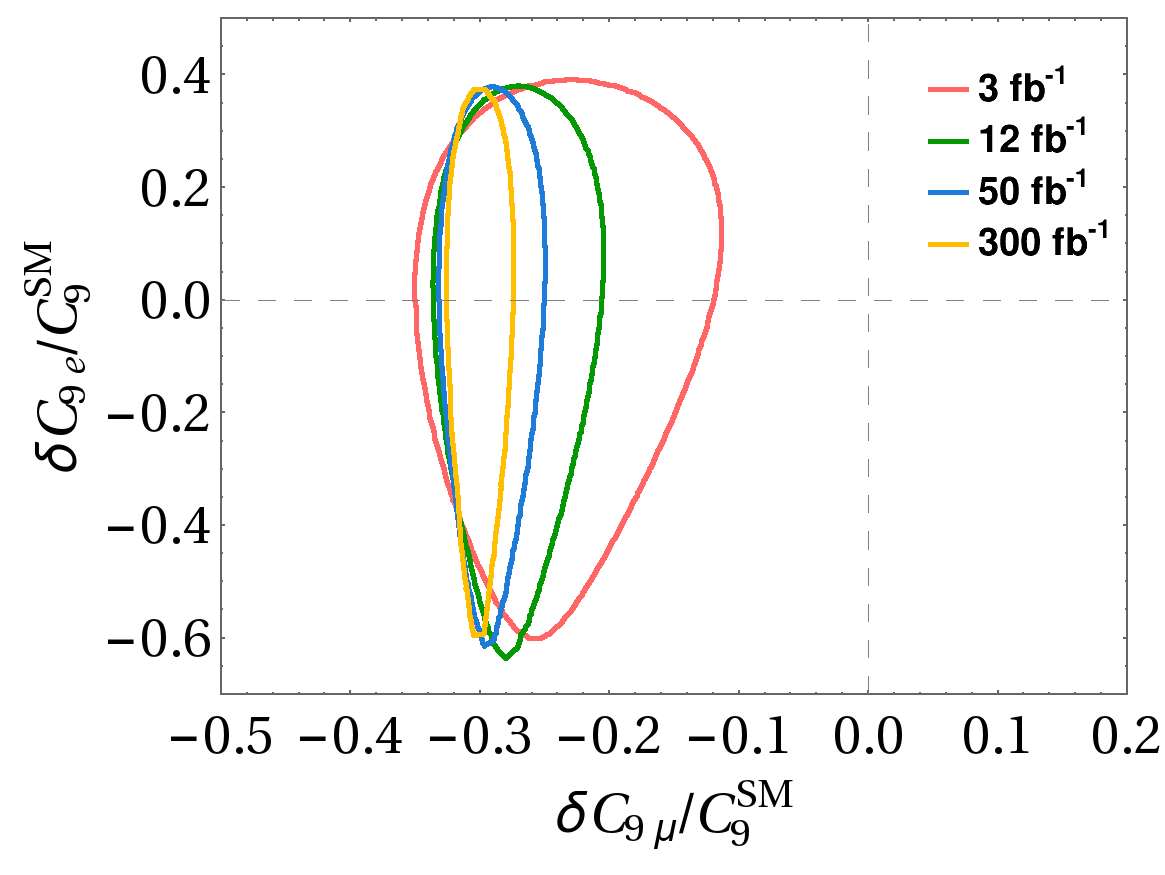}\qquad \includegraphics[width=0.47\textwidth]{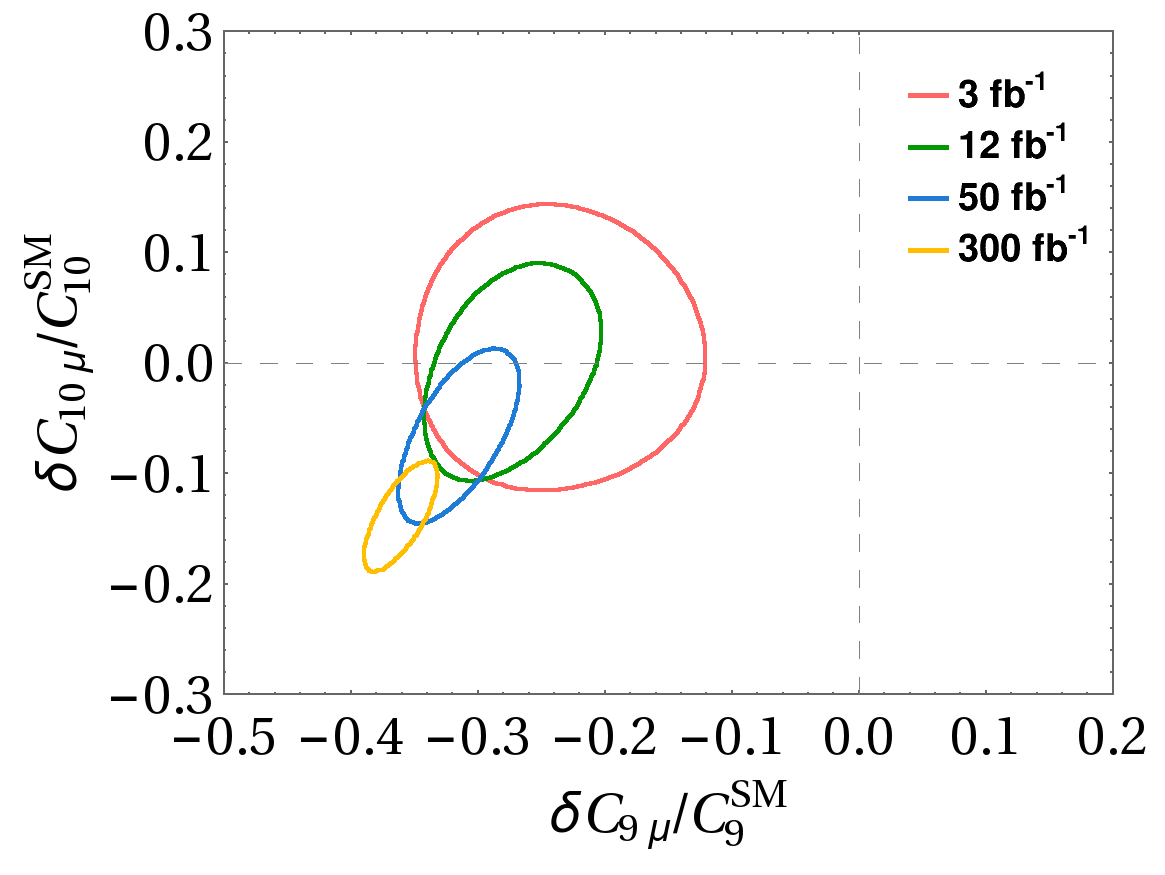}
\caption{Global fit results for $\delta C_9^e-\delta C_9^\mu$ and $\delta C_9-\delta C_{10}$, using all $b\to s \bar \ell \ell$ observables (under the assumption of 10\% factorisable power corrections) besides $R_K$ and $R_{K^*}$ are shown with a red solid line (at $2\sigma$ level).
Future LHCb prospects of the fit (at $2\sigma$ level), assuming the current central values remain, are shown with green, blue and yellow (from right to left) lines corresponding to 12, 50 and 300 fb$^{-1}$ luminosity, respectively, with the $2\sigma$ regions shrinking from right towards left.\vspace*{0.9cm}
\label{fig:2opProspect}}
\end{center}
\end{figure}

We also consider the set of $b \to s \ell\ell$ observables, which is complementary to $R_K$ and $R_K^{*}$. We again assume that  their central values  remain. Future prospects are given for two-operator fits in 
Fig.~\ref{fig:2opProspect}. Under this assumption it {seems} possible that the LHCb collaboration {will be} able to establish new physics within the angular observables even in the pessimistic case that there {will be} no theoretical progress on non-factorisable power corrections.

\vspace*{0.5cm}
\section{Predictions for other ratios based on the present measurements of $R_K$ and $R_{K^*}$}
\label{sec:predictions}
Finally, we make predictions for other ratios within the $b \to s \ell\ell$ transitions which could test lepton universality. In Ref.~\cite{Hurth:2016fbr} we made predictions based on the global fit of all $b \to s \ell\ell$ observables with $R_K$ considering 
two Wilson coefficients  $C_9^\mu$ and $C_9^e$. We found that in most cases the SM point is outside the 2$\sigma$ region of our indirect predictions reflecting the deviation in $R_K$. {Here} we base our predictions on the measurements 
of $R_K$ and $R_{K^*}$  -- assuming NP in one-operator only. We consider the six one-operator hypotheses which were favoured in our fit to the present data, see Table~\ref{tab:RxPredictions3fb}.

The predictions of the ratios are also given for the future 12 fb$^{-1}$ upgrade in Table~\ref{tab:RxPredictions12fb}, assuming the central values of the three observables $R_K^{(*)}$
remain at their current values (considering the statistical error is reduced by a factor of 2 and the systematic error remains, while no correlation among the uncertainties is assumed).

From the numbers in the last four rows of Table~\ref{tab:RxPredictions3fb} one can read off that the ratios of decay rates considered in our analysis do not help in differentiating between the six NP models. The 2$\sigma$ ranges are almost equal for all six NP options in these four cases. This will most probably not change in the future when LHCb will have collected 12 fb$^{-1}$ as one can see in our results presented in Table~\ref{tab:RxPredictions12fb}). This feature is expected when one crosschecks the analytical formulas of the decay rates (it can also be directly seen from Appendix D of Ref.~\cite{Hurth:2016fbr}).

In contrast, the ratios of the angular observables of $B \to K^* \ell\ell$ in the low-$q^2$, namely $F_L$, $A_{FB}$, and the three angular observables $S_{3}$,$S_4,$, $S_5$  are able to differentiate between the six new physics options. For example the predictions of the $2\sigma$ regions for these observables within the $C_9^\mu$ and the $C_{10}^\mu$ NP models are not overlapping in any of the cases
(see the first five rows of Table~\ref{tab:RxPredictions3fb}). And the differentiating power will increase significantly with the 12 fb$^{-1}$ data set of LHCb (see Table~ \ref{tab:RxPredictions12fb}).

However, the corresponding angular observables in the high-$q^2$ region have almost no differentiating power (see rows 7--12 in Table~\ref{tab:RxPredictions3fb}). This is expected from the well-known effect that the dependence on the Wilson coefficients and, thus, also the NP sensitivity, in general, is rather weak for observables in the high-$q^2$ region.

Some of the angular observables have zero crossings in which case it would be better to use lepton flavour differences instead of ratios~\cite{Altmannshofer:2015mqa}.
Moreover, an alternative set of observables would be the ratios and/or differences of the well-known $P_i$ observables~\cite{Capdevila:2016ivx} which are free from form factor dependences to first order. 
Predictions for all these observables are given in Tables~\ref{tab:RestPredictions3fb}~and~\ref{tab:RestPredictions12fb} in the appendix.
Further observables have been introduced where weighted differences of the angular observables are constructed~\cite{Serra:2016ivr}.

\begin{table}[t!]
\begin{center}
\rowcolors{3}{}{light-gray}
\setlength\extrarowheight{3pt}
\scalebox{0.8}{
\begin{tabular}{|l||c|c|c|c|c|c|}
\hline 
 & \multicolumn{6}{c|}{Predictions {for} 3 fb$^{-1}$ luminosity} \\ 
\hline
Obs. & $C_{9}^{\mu}$ & $C_{9}^{e}$ & $C_{10}^{\mu}$ & $C_{10}^{e}$ & $C_{LL}^{\mu}$& $C_{LL}^{e}$ \\
\hline
%
$R_{F_L}^{[1.1,6.0]}$ 	 & $ 	[0.714,0.940]	 $ &  $ 	[0.905,0.946]	 $ &  $ 	[0.996,1.059]	 $ &  $ 	[0.995,1.023]	 $ &  $ 	[0.901,0.967]	 $ &  $ 	[0.959,0.970]	$ \\
$R_{A_{FB}}^{[1.1,6.0]}$ 	 & $ 	[4.054,19.162]	 $ &  $ 	[-0.462,-0.138]	 $ &  $ 	[0.697,0.933]	 $ &  $ 	[0.954,1.099]	 $ &  $ 	[2.515,7.503]	 $ &  $ 	[-1.520,-0.212]	$ \\
$R_{S_{3}}^{[1.1,6.0]}$ 	 & $ 	[0.890,0.932]	 $ &  $ 	[0.768,0.919]	 $ &  $ 	[0.230,0.838]	 $ &  $ 	[0.714,0.873]	 $ &  $ 	[0.485,0.879]	 $ &  $ 	[0.741,0.895]	$ \\
$R_{S_{4}}^{[1.1,6.0]}$ 	 & $ 	[0.971,1.152]	 $ &  $ 	[0.822,0.950]	 $ &  $ 	[0.161,0.822]	 $ &  $ 	[0.695,0.862]	 $ &  $ 	[0.570,0.892]	 $ &  $ 	[0.755,0.903]	$ \\
$R_{S_{5}}^{[1.1,6.0]}$ 	 & $ 	[-1.450,0.637]	 $ &  $ 	[0.591,0.758]	 $ &  $ 	[0.753,1.008]	 $ &  $ 	[1.031,1.188]	 $ &  $ 	[0.293,0.833]	 $ &  $ 	[0.653,0.858]	$ \\
$R_{F_L}^{[15,19]}$ 	 & $ 	[0.999,0.998]	 $ &  $ 	[0.998,0.998]	 $ &  $ 	[0.997,0.998]	 $ &  $ 	[0.998,0.998]	 $ &  $ 	[0.998,0.998]	 $ &  $ 	[0.998,0.998]	$ \\
$R_{A_{FB}}^{[15,19]}$ 	 & $ 	[0.331,0.964]	 $ &  $ 	[0.994,1.093]	 $ &  $ 	[0.729,1.003]	 $ &  $ 	[1.027,1.169]	 $ &  $ 	[0.989,0.997]	 $ &  $ 	[0.994,0.998]	$ \\
$R_{S_{3}}^{[15,19]}$ 	 & $ 	[0.996,0.998]	 $ &  $ 	[0.998,0.999]	 $ &  $ 	[0.999,1.001]	 $ &  $ 	[0.999,1.000]	 $ &  $ 	[0.999,1.000]	 $ &  $ 	[0.998,0.999]	$ \\
$R_{S_{4}}^{[15,19]}$ 	 & $ 	[0.998,0.999]	 $ &  $ 	[0.998,0.998]	 $ &  $ 	[0.998,0.998]	 $ &  $ 	[0.998,0.999]	 $ &  $ 	[0.998,0.999]	 $ &  $ 	[0.998,0.998]	$ \\
$R_{S_{5}}^{[15,19]}$ 	 & $ 	[0.330,0.964]	 $ &  $ 	[0.994,1.092]	 $ &  $ 	[0.729,1.003]	 $ &  $ 	[1.027,1.169]	 $ &  $ 	[0.989,0.997]	 $ &  $ 	[0.994,0.997]	$ \\
$R_{K^*}^{[15,19]}$ 	 & $ 	[0.572,0.867]	 $ &  $ 	[0.520,0.841]	 $ &  $ 	[0.529,0.844]	 $ &  $ 	[0.530,0.838]	 $ &  $ 	[0.523,0.847]	 $ &  $ 	[0.525,0.839]	$ \\
$R_{K}^{[15,19]}$ 	 & $ 	[0.527,0.869]	 $ &  $ 	[0.534,0.846]	 $ &  $ 	[0.610,0.873]	 $ &  $ 	[0.532,0.908]	 $ &  $ 	[0.561,0.863]	 $ &  $ 	[0.554,0.855]	$ \\
$R_{\phi}^{[1.1,6.0]}$ 	 & $ 	[0.740,0.897]	 $ &  $ 	[0.561,0.867]	 $ &  $ 	[0.517,0.838]	 $ &  $ 	[0.476,0.883]	 $ &  $ 	[0.576,0.860]	 $ &  $ 	[0.543,0.849]	$ \\
$R_{\phi}^{[15,19]}$ 	 & $ 	[0.575,0.867]	 $ &  $ 	[0.520,0.841]	 $ &  $ 	[0.526,0.843]	 $ &  $ 	[0.481,0.887]	 $ &  $ 	[0.521,0.847]	 $ &  $ 	[0.524,0.839]	$ \\
\hline
\end{tabular}
}
\caption{
Predictions of ratios of observables with muons in the final state to electrons in the final
state at 95\% confidence level, considering one-operator fits obtained using the 3 fb$^{-1}$ data for $R_{K^{(*)}}$.
The observables $R_{F_{L}}, R_{A_{FB}}, R_{S_{3,4,5}}$ correspond to ratios of $F_{L}, A_{FB},S_{3,4,5}$ of the $B\to K^* \bar \ell \ell$ decay, respectively. 
The observables $R_{K^{(*)}}$ and $R_\phi$ correspond to the ratios of the branching fractions of $B\to K^{(*)} \bar \ell \ell$
and $B_s\to \phi \bar \ell \ell$, respectively. 
The superscripts denote the $q^2$ bins.
\label{tab:RxPredictions3fb}} 
\end{center} 
\end{table}
\begin{table}
\begin{center}
\rowcolors{3}{}{light-gray}
\setlength\extrarowheight{3pt}
\scalebox{0.8}{
\begin{tabular}{|l||c|c|c|c|c|c|}
\hline 
 & \multicolumn{6}{c|}{Predictions assuming 12 fb$^{-1}$ luminosity} \\ 
\hline
Obs. & $C_{9}^{\mu}$ & $C_{9}^{e}$ & $C_{10}^{\mu}$ & $C_{10}^{e}$ & $C_{LL}^{\mu}$& $C_{LL}^{e}$ \\
\hline
$R_{F_L}^{[1.1,6.0]}$ 	 & $ 	[0.785,0.913]	 $ &  $ 	[0.909,0.933]	 $ &  $ 	[1.005,1.042]	 $ &  $ 	[1.001,1.018]	 $ &  $ 	[0.920,0.958]	 $ &  $ 	[0.960,0.966]	$ \\
$R_{A_{FB}}^{[1.1,6.0]}$ & $ 	[6.048,14.819]	 $ &  $ 	[-0.288,-0.153]	 $ &  $ 	[0.816,0.928]	 $ &  $ 	[0.974,1.061]	 $ &  $ 	[3.338,6.312]	 $ &  $ 	[-0.684,-0.256]	$ \\
$R_{S_{3}}^{[1.1,6.0]}$  & $ 	[0.858,0.904]	 $ &  $ 	[0.795,0.886]	 $ &  $ 	[0.399,0.753]	 $ &  $ 	[0.738,0.832]	 $ &  $ 	[0.586,0.819]	 $ &  $ 	[0.766,0.858]	$ \\
$R_{S_{4}}^{[1.1,6.0]}$  & $ 	[0.970,1.051]	 $ &  $ 	[0.848,0.926]	 $ &  $ 	[0.344,0.730]	 $ &  $ 	[0.719,0.818]	 $ &  $ 	[0.650,0.841]	 $ &  $ 	[0.780,0.868]	$ \\
$R_{S_{5}}^{[1.1,6.0]}$  & $ 	[-0.787,0.394]	 $ &  $ 	[0.603,0.697]	 $ &  $ 	[0.881,1.002]	 $ &  $ 	[1.053,1.146]	 $ &  $ 	[0.425,0.746]	 $ &  $ 	[0.685,0.806]	$ \\
$R_{F_L}^{[15,19]}$ 	 & $ 	[0.999,0.999]	 $ &  $ 	[0.998,0.998]	 $ &  $ 	[0.997,0.998]	 $ &  $ 	[0.998,0.998]	 $ &  $ 	[0.998,0.998]	 $ &  $ 	[0.998,0.998]	$ \\
$R_{A_{FB}}^{[15,19]}$ 	 & $ 	[0.616,0.927]	 $ &  $ 	[1.002,1.061]	 $ &  $ 	[0.860,0.994]	 $ &  $ 	[1.046,1.131]	 $ &  $ 	[0.992,0.996]	 $ &  $ 	[0.995,0.997]	$ \\
$R_{S_{3}}^{[15,19]}$ 	 & $ 	[0.997,0.998]	 $ &  $ 	[0.998,0.998]	 $ &  $ 	[0.999,1.000]	 $ &  $ 	[0.999,1.000]	 $ &  $ 	[0.999,1.000]	 $ &  $ 	[0.999,0.999]	$ \\
$R_{S_{4}}^{[15,19]}$ 	 & $ 	[0.998,0.999]	 $ &  $ 	[0.998,0.998]	 $ &  $ 	[0.998,0.998]	 $ &  $ 	[0.998,0.999]	 $ &  $ 	[0.998,0.998]	 $ &  $ 	[0.998,0.998]	$ \\
$R_{S_{5}}^{[15,19]}$ 	 & $ 	[0.615,0.927]	 $ &  $ 	[1.002,1.061]	 $ &  $ 	[0.860,0.994]	 $ &  $ 	[1.046,1.131]	 $ &  $ 	[0.991,0.996]	 $ &  $ 	[0.994,0.997]	$ \\
$R_{K^*}^{[15,19]}$ 	 & $ 	[0.621,0.803]	 $ &  $ 	[0.577,0.771]	 $ &  $ 	[0.589,0.778]	 $ &  $ 	[0.586,0.770]	 $ &  $ 	[0.585,0.780]	 $ &  $ 	[0.582,0.771]	$ \\
$R_{K}^{[15,19]}$ 	 & $ 	[0.597,0.802]	 $ &  $ 	[0.590,0.778]	 $ &  $ 	[0.659,0.818]	 $ &  $ 	[0.632,0.805]	 $ &  $ 	[0.620,0.802]	 $ &  $ 	[0.609,0.791]	$ \\
$R_{\phi}^{[1.1,6.0]}$ 	 & $ 	[0.748,0.852]	 $ &  $ 	[0.620,0.805]	 $ &  $ 	[0.578,0.770]	 $ &  $ 	[0.578,0.764]	 $ &  $ 	[0.629,0.800]	 $ &  $ 	[0.600,0.784]	$ \\
$R_{\phi}^{[15,19]}$ 	 & $ 	[0.623,0.803]	 $ &  $ 	[0.577,0.771]	 $ &  $ 	[0.586,0.776]	 $ &  $ 	[0.583,0.769]	 $ &  $ 	[0.584,0.779]	 $ &  $ 	[0.581,0.770]	$ \\
\hline
\end{tabular}
}
\caption{
Predictions of ratios of observables with muons in the final state to electrons in the final
state at 95\% confidence level, considering one-operator fits obtained by assuming the central values of $R_{K^{(*)}}$ with 12 fb$^{-1}$ luminosity remain the same as the current 3 fb$^{-1}$ data.
In a few cases, the 12 fb$^{-1}$ predictions are not fully within the 3 fb$^{-1}$ prediction ranges which is due to a change in the position of the minimum.
For the definition of the observables see the caption of Table~\ref{tab:RxPredictions3fb}.
\label{tab:RxPredictions12fb}} 
\end{center} 
\end{table}

\vspace*{0.5cm}
\section{Conclusions}
The future measurements of the theoretically clean ratios $R_{K^{(*)}}$  and similar observables which 
are sensitive to lepton flavour non-universality have the potential to unambiguously establish lepton non-universal new physics in the near future. We have demonstrated that this {may} be possible already with the 12 fb$^{-1}$ data set of LHCb. We have also shown that more such theoretically clean ratios are needed to differentiate between the six NP hypotheses favoured by the present data. We have singled out the ratios of the angular observables of the $B \to K^* \ell\ell$ decay in the low $q^2$ region to have the largest differentiating power in this respect.
Such a finding of lepton-flavour non-universal new physics  may also indirectly establish the new physics explanation of the present anomalies in the angular observables in $B \to K^{*} \mu\mu$ decays and in the branching ratios of $B_s \to \phi \mu\mu$ if there is a coherent NP picture of both sets of observables.

\clearpage

\appendix

\section{Predictions for further observables sensitive to lepton flavour violation} 

\begin{table}[h!]
\begin{center}
\vspace*{1.cm}
\rowcolors{3}{}{light-gray}
\setlength\extrarowheight{3pt}
\scalebox{0.75}{
\begin{tabular}{|l||c|c|c|c|c|c|}
\hline 
 & \multicolumn{6}{c|}{Predictions for 3 fb$^{-1}$ luminosity} \\ 
\hline
Obs. & $C_{9}^{\mu}$ & $C_{9}^{e}$ & $C_{10}^{\mu}$ & $C_{10}^{e}$ & $C_{LL}^{\mu}$& $C_{LL}^{e}$ \\
\hline
$D_{F_L}^{[1.1,6.0]}$ 	 & $ 	[-0.046,-0.218]	 $ &  $ 	[-0.078,-0.043]	 $ &  $ 	[-0.003,0.045]	 $ &  $ 	[-0.004,0.017]	 $ &  $ 	[-0.076,-0.025]	 $ &  $ 	[-0.032,-0.023]	$ \\
$D_{A_{FB}}^{[1.1,6.0]}$ 	 & $ 	[-0.247,-0.041]	 $ &  $ 	[-0.104,-0.040]	 $ &  $ 	[0.001,0.004]	 $ &  $ 	[-0.001,0.001]	 $ &  $ 	[-0.088,-0.021]	 $ &  $ 	[-0.072,-0.021]	$ \\
$D_{S_{3}}^{[1.1,6.0]}$ 	 & $ 	[0.001,0.001]	 $ &  $ 	[0.001,0.003]	 $ &  $ 	[0.002,0.009]	 $ &  $ 	[0.002,0.005]	 $ &  $ 	[0.001,0.006]	 $ &  $ 	[0.001,0.004]	$ \\
$D_{S_{4}}^{[1.1,6.0]}$ 	 & $ 	[-0.004,0.019]	 $ &  $ 	[-0.027,-0.007]	 $ &  $ 	[-0.106,-0.023]	 $ &  $ 	[-0.055,-0.020]	 $ &  $ 	[-0.055,-0.014]	 $ &  $ 	[-0.041,-0.014]	$ \\
$D_{S_{5}}^{[1.1,6.0]}$ 	 & $ 	[0.059,0.401]	 $ &  $ 	[0.052,0.113]	 $ &  $ 	[-0.001,0.040]	 $ &  $ 	[-0.026,-0.005]	 $ &  $ 	[0.027,0.116]	 $ &  $ 	[0.027,0.087]	$ \\
$D_{F_L}^{[15,19]}$ 	 & $ 	[-0.001,-0.000]	 $ &  $ 	[-0.001,-0.001]	 $ &  $ 	[-0.001,-0.001]	 $ &  $ 	[-0.001,-0.001]	 $ &  $ 	[-0.001,-0.001]	 $ &  $ 	[-0.001,-0.001]	$ \\
$D_{A_{FB}}^{[15,19]}$ 	 & $ 	[-0.254,-0.014]	 $ &  $ 	[-0.002,0.032]	 $ &  $ 	[-0.103,0.001]	 $ &  $ 	[0.010,0.055]	 $ &  $ 	[-0.004,-0.001]	 $ &  $ 	[-0.002,-0.001]	$ \\
$D_{S_{3}}^{[15,19]}$ 	 & $ 	[0.000,0.001]	 $ &  $ 	[0.000,0.000]	 $ &  $ 	[-0.000,0.000]	 $ &  $ 	[-0.000,0.000]	 $ &  $ 	[0.000,0.000]	 $ &  $ 	[0.000,0.000]	$ \\
$D_{S_{4}}^{[15,19]}$ 	 & $ 	[-0.001,-0.000]	 $ &  $ 	[-0.001,-0.001]	 $ &  $ 	[-0.001,-0.001]	 $ &  $ 	[-0.001,-0.000]	 $ &  $ 	[-0.001,-0.000]	 $ &  $ 	[-0.001,-0.000]	$ \\
$D_{S_{5}}^{[15,19]}$ 	 & $ 	[0.010,0.191]	 $ &  $ 	[-0.024,0.002]	 $ &  $ 	[-0.001,0.077]	 $ &  $ 	[-0.041,-0.007]	 $ &  $ 	[0.001,0.003]	 $ &  $ 	[0.001,0.002]	$ \\
\hline
$R_{P_{2}}^{[1.1,6.0]}$ 	 & $ 	[3.653,10.426]	 $ &  $ 	[-0.416,-0.103]	 $ &  $ 	[0.983,1.009]	 $ &  $ 	[1.019,1.272]	 $ &  $ 	[2.461,6.105]	 $ &  $ 	[-1.499,-0.201]	$ \\
$R_{P_{1}}^{[1.1,6.0]}$ 	 & $ 	[0.484,0.840]	 $ &  $ 	[0.575,0.827]	 $ &  $ 	[0.325,0.906]	 $ &  $ 	[0.826,0.933]	 $ &  $ 	[0.395,0.860]	 $ &  $ 	[0.702,0.882]	$ \\
$R_{P_{4}}^{[1.1,6.0]}$ 	 & $ 	[0.951,1.006]	 $ &  $ 	[0.747,0.927]	 $ &  $ 	[0.186,0.857]	 $ &  $ 	[0.739,0.894]	 $ &  $ 	[0.541,0.898]	 $ &  $ 	[0.750,0.910]	$ \\
$R_{P_{5}}^{[1.1,6.0]}$ 	 & $ 	[-1.266,0.624]	 $ &  $ 	[0.537,0.740]	 $ &  $ 	[0.869,1.050]	 $ &  $ 	[1.068,1.263]	 $ &  $ 	[0.278,0.838]	 $ &  $ 	[0.649,0.865]	$ \\
$R_{P_{2}}^{[15,19]}$ 	 & $ 	[0.331,0.966]	 $ &  $ 	[0.996,1.095]	 $ &  $ 	[0.732,1.006]	 $ &  $ 	[1.029,1.171]	 $ &  $ 	[0.991,0.999]	 $ &  $ 	[0.996,1.000]	$ \\
$R_{P_{1}}^{[15,19]}$ 	 & $ 	[0.997,1.000]	 $ &  $ 	[1.000,1.001]	 $ &  $ 	[1.001,1.005]	 $ &  $ 	[1.001,1.003]	 $ &  $ 	[1.001,1.002]	 $ &  $ 	[1.001,1.001]	$ \\
$R_{P_{4}}^{[15,19]}$ 	 & $ 	[0.999,1.000]	 $ &  $ 	[1.000,1.000]	 $ &  $ 	[1.000,1.001]	 $ &  $ 	[1.000,1.001]	 $ &  $ 	[1.000,1.000]	 $ &  $ 	[1.000,1.000]	$ \\
$R_{P_{5}}^{[15,19]}$ 	 & $ 	[0.330,0.966]	 $ &  $ 	[0.996,1.094]	 $ &  $ 	[0.732,1.006]	 $ &  $ 	[1.029,1.171]	 $ &  $ 	[0.991,0.999]	 $ &  $ 	[0.996,0.999]	$ \\
\hline
$Q_{2}^{[1.1,6.0]}$ 	 & $ 	[0.102,0.361]	 $ &  $ 	[0.123,0.387]	 $ &  $ 	[-0.001,0.000]	 $ &  $ 	[0.001,0.008]	 $ &  $ 	[0.056,0.196]	 $ &  $ 	[0.060,0.216]	$ \\
$Q_{1}^{[1.1,6.0]}$ 	 & $ 	[0.016,0.050]	 $ &  $ 	[0.021,0.073]	 $ &  $ 	[0.009,0.066]	 $ &  $ 	[0.007,0.021]	 $ &  $ 	[0.014,0.059]	 $ &  $ 	[0.013,0.042]	$ \\
$Q_{4}^{[1.1,6.0]}$ 	 & $ 	[-0.029,0.003]	 $ &  $ 	[-0.204,-0.048]	 $ &  $ 	[-0.486,-0.086]	 $ &  $ 	[-0.213,-0.072]	 $ &  $ 	[-0.274,-0.061]	 $ &  $ 	[-0.201,-0.060]	$ \\
$Q_{5}^{[1.1,6.0]}$ 	 & $ 	[0.145,0.872]	 $ &  $ 	[0.138,0.338]	 $ &  $ 	[-0.019,0.051]	 $ &  $ 	[-0.082,-0.025]	 $ &  $ 	[0.062,0.278]	 $ &  $ 	[0.061,0.212]	$ \\
$Q_{2}^{[15,19]}$ 	 & $ 	[0.013,0.254]	 $ &  $ 	[-0.033,0.001]	 $ &  $ 	[-0.002,0.102]	 $ &  $ 	[-0.056,-0.011]	 $ &  $ 	[0.000,0.003]	 $ &  $ 	[0.000,0.002]	$ \\
$Q_{1}^{[15,19]}$ 	 & $ 	[0.000,0.002]	 $ &  $ 	[-0.000,-0.000]	 $ &  $ 	[-0.003,-0.001]	 $ &  $ 	[-0.002,-0.001]	 $ &  $ 	[-0.001,-0.000]	 $ &  $ 	[-0.001,-0.000]	$ \\
$Q_{4}^{[15,19]}$ 	 & $ 	[-0.001,-0.000]	 $ &  $ 	[0.000,0.000]	 $ &  $ 	[0.000,0.001]	 $ &  $ 	[0.000,0.001]	 $ &  $ 	[0.000,0.000]	 $ &  $ 	[0.000,0.000]	$ \\
$Q_{5}^{[15,19]}$ 	 & $ 	[0.021,0.405]	 $ &  $ 	[-0.052,0.002]	 $ &  $ 	[-0.003,0.162]	 $ &  $ 	[-0.088,-0.017]	 $ &  $ 	[0.000,0.005]	 $ &  $ 	[0.000,0.003]	$ \\
\hline
\end{tabular}
}
\caption{
Predictions of ratios and differences of observables with muons in the final state to electrons in the final
state at 95\% confidence level, considering one-operator fits obtained {using} the 3 fb$^{-1}$ data for $R_{K^{(*)}}$.
The observables $D_{S_{3,4,5}} = S_{3,4,5}^{\mu}-S_{3,4,5}^{e}$, $D_{A_{FB}} = A_{FB}^{\mu}-A_{FB}^{e}$, $D_{F_L} = Q_{F_L} = F_{L}^{\mu}-F_{L}^{e}$ 
and $Q_{1,2,4,5} = P_{1,2,4,5}^{(\prime)\;\mu}-P_{1,2,4,5}^{(\prime)\;e}$ and $R_{P_i} = P_{1,2,4,5}^{(\prime)\;\mu} /P_{1,2,4,5}^{(\prime)\;e} $
all correspond to the  $B\to K^* \bar \ell \ell$ decay. 
The observables $R_{K^{(*)}},R_{X_s}$ and $R_\phi$ correspond to the ratios of the branching fractions of $B\to K^{(*)} \bar \ell \ell,B \to X_s \bar \ell \ell$
and $B_s\to \phi \bar \ell \ell$, respectively. 
The superscripts  denote the $q^2$ bins.
\label{tab:RestPredictions3fb}} 
\end{center} 
\end{table}

\begin{table}
\begin{center}
\rowcolors{3}{}{light-gray}
\setlength\extrarowheight{3pt}
\scalebox{0.75}{
\begin{tabular}{|l||c|c|c|c|c|c|}
\hline 
 & \multicolumn{6}{c|}{Predictions assuming 12 fb$^{-1}$ luminosity} \\ 
\hline
Obs. & $C_{9}^{\mu}$ & $C_{9}^{e}$ & $C_{10}^{\mu}$ & $C_{10}^{e}$ & $C_{LL}^{\mu}$& $C_{LL}^{e}$ \\
\hline
$D_{F_L}^{[1.1,6.0]}$ 	 & $	[-0.164,-0.066]	 $ &  $ 	[-0.075,-0.053]	 $ &  $ 	[0.004,0.032]	 $ &  $ 	[0.000,0.013]	 $ &  $ 	[-0.061,-0.032]	 $ &  $ 	[-0.032,-0.026]	$ \\
$D_{A_{FB}}^{[1.1,6.0]}$ & $ 	[-0.188,-0.069]	 $ &  $ 	[-0.095,-0.056]	 $ &  $ 	[0.001,0.003]	 $ &  $ 	[-0.001,0.000]	 $ &  $ 	[-0.072,-0.032]	 $ &  $ 	[-0.062,-0.031]	$ \\
$D_{S_{3}}^{[1.1,6.0]}$  & $	 [0.001,0.002]	 $ &  $ 	[0.001,0.003]	 $ &  $ 	[0.003,0.007]	 $ &  $ 	[0.002,0.004]	 $ &  $ 	[0.002,0.005]	 $ &  $ 	[0.002,0.004]	$ \\
$D_{S_{4}}^{[1.1,6.0]}$  & $ 	[-0.004,0.007]	 $ &  $ 	[-0.023,-0.010]	 $ &  $ 	[-0.083,-0.034]	 $ &  $ 	[-0.049,-0.028]	 $ &  $ 	[-0.044,-0.020]	 $ &  $ 	[-0.035,-0.019]	$ \\
$D_{S_{5}}^{[1.1,6.0]}$  & $ 	[0.099,0.292]	 $ &  $ 	[0.071,0.107]	 $ &  $ 	[-0.000,0.019]	 $ &  $ 	[-0.021,-0.008]	 $ &  $ 	[0.041,0.094]	 $ &  $ 	[0.039,0.075]	$ \\
$D_{F_L}^{[15,19]}$ 	 & $ 	[-0.001,-0.000]	 $ &  $ 	[-0.001,-0.001]	 $ &  $ 	[-0.001,-0.001]	 $ &  $ 	[-0.001,-0.001]	 $ &  $ 	[-0.001,-0.001]	 $ &  $ 	[-0.001,-0.001]	$ \\
$D_{A_{FB}}^{[15,19]}$ 	 & $ 	[-0.146,-0.027]	 $ &  $ 	[0.001,0.022]	 $ &  $ 	[-0.053,-0.002]	 $ &  $ 	[0.017,0.044]	 $ &  $ 	[-0.003,-0.001]	 $ &  $ 	[-0.002,-0.001]	$ \\
$D_{S_{3}}^{[15,19]}$ 	 & $ 	[0.000,0.001]	 $ &  $ 	[0.000,0.000]	 $ &  $ 	[-0.000,0.000]	 $ &  $ 	[-0.000,0.000]	 $ &  $ 	[0.000,0.000]	 $ &  $ 	[0.000,0.000]	$ \\
$D_{S_{4}}^{[15,19]}$ 	 & $ 	[-0.000,-0.000]	 $ &  $ 	[-0.001,-0.001]	 $ &  $ 	[-0.001,-0.001]	 $ &  $ 	[-0.001,-0.000]	 $ &  $ 	[-0.001,-0.000]	 $ &  $ 	[-0.001,-0.001]	$ \\
$D_{S_{5}}^{[15,19]}$ 	 & $ 	[0.021,0.110]	 $ &  $ 	[-0.016,-0.001]	 $ &  $ 	[0.002,0.040]	 $ &  $ 	[-0.033,-0.012]	 $ &  $ 	[0.001,0.002]	 $ &  $ 	[0.001,0.002]	$ \\
\hline
$R_{P_{2}}^{[1.1,6.0]}$  & $ 	[5.054,9.201]	 $ &  $ 	[-0.246,-0.117]	 $ &  $ 	[1.037,1.061]	 $ &  $ 	[1.060,1.211]	 $ &  $ 	[3.185,5.402]	 $ &  $ 	[-0.666,-0.243]	$ \\
$R_{P_{1}}^{[1.1,6.0]}$  & $ 	[0.532,0.755]	 $ &  $ 	[0.607,0.758]	 $ &  $ 	[0.518,0.842]	 $ &  $ 	[0.843,0.906]	 $ &  $ 	[0.501,0.781]	 $ &  $ 	[0.727,0.835]	$ \\
$R_{P_{4}}^{[1.1,6.0]}$  & $ 	[0.928,0.935]	 $ &  $ 	[0.777,0.887]	 $ &  $ 	[0.385,0.770]	 $ &  $ 	[0.762,0.854]	 $ &  $ 	[0.626,0.839]	 $ &  $ 	[0.776,0.871]	$ \\
$R_{P_{5}}^{[1.1,6.0]}$  & $ 	[-0.700,0.377]	 $ &  $ 	[0.553,0.668]	 $ &  $ 	[0.985,1.057]	 $ &  $ 	[1.098,1.214]	 $ &  $ 	[0.410,0.745]	 $ &  $ 	[0.681,0.810]	$ \\
$R_{P_{2}}^{[15,19]}$ 	 & $ 	[0.616,0.929]	 $ &  $ 	[1.004,1.063]	 $ &  $ 	[0.863,0.996]	 $ &  $ 	[1.048,1.133]	 $ &  $ 	[0.994,0.998]	 $ &  $ 	[0.997,0.999]	$ \\
$R_{P_{1}}^{[15,19]}$ 	 & $ 	[0.998,1.000]	 $ &  $ 	[1.000,1.000]	 $ &  $ 	[1.002,1.004]	 $ &  $ 	[1.001,1.002]	 $ &  $ 	[1.001,1.002]	 $ &  $ 	[1.001,1.001]	$ \\
$R_{P_{4}}^{[15,19]}$ 	 & $ 	[1.000,1.000]	 $ &  $ 	[1.000,1.000]	 $ &  $ 	[1.000,1.001]	 $ &  $ 	[1.000,1.000]	 $ &  $ 	[1.000,1.000]	 $ &  $ 	[1.000,1.000]	$ \\
$R_{P_{5}}^{[15,19]}$ 	 & $ 	[0.615,0.929]	 $ &  $ 	[1.004,1.063]	 $ &  $ 	[0.863,0.996]	 $ &  $ 	[1.048,1.133]	 $ &  $ 	[0.993,0.998]	 $ &  $ 	[0.996,0.999]	$ \\
\hline
$Q_{2}^{[1.1,6.0]}$ 	 & $ 	[0.155,0.314]	 $ &  $ 	[0.183,0.346]	 $ &  $ 	[0.001,0.002]	 $ &  $ 	[0.002,0.006]	 $ &  $ 	[0.084,0.169]	 $ &  $ 	[0.090,0.185]	$ \\
$Q_{1}^{[1.1,6.0]}$ 	 & $ 	[0.024,0.046]	 $ &  $ 	[0.032,0.064]	 $ &  $ 	[0.015,0.047]	 $ &  $ 	[0.010,0.019]	 $ &  $ 	[0.021,0.049]	 $ &  $ 	[0.020,0.037]	$ \\
$Q_{4}^{[1.1,6.0]}$ 	 & $ 	[-0.043,-0.039]	 $ &  $ 	[-0.173,-0.077]	 $ &  $ 	[-0.367,-0.137]	 $ &  $ 	[-0.188,-0.104]	 $ &  $ 	[-0.223,-0.096]	 $ &  $ 	[-0.174,-0.089]	$ \\
$Q_{5}^{[1.1,6.0]}$ 	 & $ 	[0.240,0.654]	 $ &  $ 	[0.195,0.317]	 $ &  $ 	[-0.022,0.006]	 $ &  $ 	[-0.069,-0.035]	 $ &  $ 	[0.098,0.227]	 $ &  $ 	[0.092,0.183]	$ \\
$Q_{2}^{[15,19]}$ 	 & $ 	[0.027,0.146]	 $ &  $ 	[-0.023,-0.001]	 $ &  $ 	[0.001,0.052]	 $ &  $ 	[-0.045,-0.017]	 $ &  $ 	[0.001,0.002]	 $ &  $ 	[0.000,0.001]	$ \\
$Q_{1}^{[15,19]}$ 	 & $ 	[0.000,0.001]	 $ &  $ 	[-0.000,-0.000]	 $ &  $ 	[-0.002,-0.001]	 $ &  $ 	[-0.001,-0.001]	 $ &  $ 	[-0.001,-0.000]	 $ &  $ 	[-0.001,-0.000]	$ \\
$Q_{4}^{[15,19]}$ 	 & $ 	[-0.000,-0.000]	 $ &  $ 	[0.000,0.000]	 $ &  $ 	[0.000,0.001]	 $ &  $ 	[0.000,0.001]	 $ &  $ 	[0.000,0.000]	 $ &  $ 	[0.000,0.000]	$ \\
$Q_{5}^{[15,19]}$ 	 & $ 	[0.232,0.043]	 $ &  $ 	[-0.036,-0.002]	 $ &  $ 	[0.002,0.083]	 $ &  $ 	[-0.071,-0.028]	 $ &  $ 	[0.001,0.004]	 $ &  $ 	[0.001,0.002]	$ \\
\hline
\end{tabular}
}
\caption{
Predictions of  ratios and differences of observables with muons in the final state to electrons in the final state at 95\% confidence level, considering one-operator fits  obtained by assuming the central values of $R_{K^{(*)}}$ with 12 fb$^{-1}$ luminosity remain the same as the current 3 fb$^{-1}$ data.
For the definition of the observables see the caption of Table~\ref{tab:RestPredictions3fb}.
\label{tab:RestPredictions12fb}} 
\end{center} 
\end{table}

\newpage\newpage

\providecommand{\href}[2]{#2}\begingroup\raggedright\endgroup

\end{document}